\documentstyle[aps,prl,twocolumn]{revtex}
\begin{document}
\draft
\wideabs{
\title{Magnetic Polarization Currents in Double Quantum Dot Devices}
\author{Sam Young Cho$^{1}$, Ross H. McKenzie$^1$,
        Kicheon Kang$^2$, and Chul Koo Kim$^{3}$}
\address{$^1$Department of Physics, University of Queensland,
             Brisbane 4072, Australia \\
         $^2$Basic Research Laboratory, 
             Electronics and Telecommunications Research Institute, 
             Taejon 305-350, Korea \\
         $^3$Institute of Physics and
             Applied Physics, Yonsei University, Seoul 120-749, Korea 
}
\date{\today}
\maketitle

\begin{abstract}
 We investigate coherent electron transport through 
 a parallel circuit of
 two quantum dots, each of which has a
 single tunable energy level.
 Electrons tunneling via each dot from the left lead 
 interfere with each other at the right lead.
 It is shown that due to the quantum interference of tunneling electrons
 the double quantum dot device 
 is magnetically polarized by coherent circulation of 
 electrons 
 on the closed path through the dots and the leads.
 Varying the energy level of each dot one can make  
 the magnetic states of the device to be either   
 {\em up-}, {\em non-}, or {\em down-} polarization.
 It is shown that for experimentally accessible
 temperatures and applied biases
 the magnetic polarization currents should be sufficiently
 large to observe with current nanotechnology.
\end{abstract}
%
%
%
}

 The most interesting
 phenomena seen in mesoscopic electronic devices
 are due to the
 quantum coherence of electrons being 
 maintained over a significant part of the transport process.
 Examples of such interference effects
 \cite{review}
 that have been observed include
 weak localization, 
 universal conductance fluctuations,
 and Aharonov-Bohm (AB) oscillations.
 In 1995, Yacoby and coworkers\cite{Yacobi95}
 demonstrated the coherence of electron waves passing by
 resonant tunneling through a quantum dot (QD) in a double-slit type
 interference in a ring geometry.
 Recent interference experiments
 \cite{Schuster97,Wiel00,Ji00} 
 with two different transport paths 
 in a ring geometry have enabled the realization of 
 a phase sensitive probe of the effects of 
 electron-electron interaction on the conductance oscillation
 such as Kondo correlations\cite{Ji00,Gerland00,Hofstetter01,Kang00,Cho01},
 as well as the anomalous phase of the transmission coefficients
 through a QD  
 \cite{Yeyati95,Hacken96,Bruder96,Cho98,Lee99,Baltin99}.
 Two QDs have also been fabricated 
 experimentally on two different electron pathways
 \cite{Holleitner00}.
 These double quantum dot (DQD) devices
 provide a good opportunity to test theories 
 of resonant tunneling \cite{Shahbazyan94,Bjorn01}, 
 cotunneling \cite{Akera93,Loss00,Koenig01},
 and many-body correlation effects\cite{Boese01,Hofstetter01.2}.
 Compared to ballistic electron interference devices\cite{Mailly93,Rabaud01},
 a DQD device makes it possible to manipulate 
 the coherent tunneling of electrons through each dot 
 separately by varying the gate voltages of the dots.
 K\"onig and Gefen\cite{Koenig01}
 have discussed quantum coherence
 in DQD devices with the {\em same} energy level in each dot.

\begin{figure}
\vspace*{3.5cm}
\includegraphics{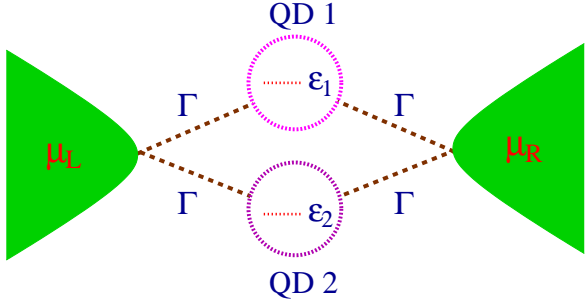}
\caption{A double quantum dot (DQD) device. 
  Both dots are tunnel-coupled to 
  the left and right leads.
  The leads are characterized by
  the chemical potentials, $\mu_L$ and $\mu_R$.
  The tunneling amplitudes between the dots and the leads
  are denoted by $\Gamma$.
  The energy level position in each dot is measured 
  as $\varepsilon_1$ and $\varepsilon_2$ from the Fermi energy
  in the leads.
  }
\label{fig1}
\end{figure}

 In this Letter, we study coherent electron transport
 through two parallel QDs, each of which
 has a  single {\it tunable} energy level
 (see Fig. \ref{fig1}).
 Remarkably, we find a coherent magnetic polarization current (MPC)
 circulating
 on the closed path connecting the dots and the leads
 as a function of each dot level position. 
 This MPC 
 is induced by coherent tunneling
 for electron transport through each QD.
 We discuss the magnetic polarizability of the DQD device
 due to the MPC 
 for finite temperature and finite applied bias.
 
 We start with the model Hamiltonian
\begin{equation}
  H\!\! = \!\!\!\!\! \sum_{k\in L,R \atop \sigma} \!\!\!\!
        \varepsilon_{k} c^\dagger_{k\sigma}c_{k\sigma} \!\!
     + \!\!\! \sum_{i\in 1,2 \atop \sigma} \!\!\!
        \varepsilon_i d^\dagger_{i\sigma} d_{i\sigma} \!\!
     + \!\!\!\!\! \sum_{i\in 1,2 \atop k\sigma\in L,R} \!\!\! 
        (V_{k} c^\dagger_{k\sigma} d_{i\sigma}\!\! +\! {\rm h.c.}) ,
\label{Hamiltonian}
\end{equation}
 where $c_{k\sigma}$ and $d_{i\sigma}$ are the annihilation operators 
 with spin $\sigma$ 
 for electrons in the leads and the dots $(i=1,2)$, respectively.
 $\varepsilon_1$ and $\varepsilon_2$ are the level energy in each dot,
 measured, relative to the Fermi energy of the leads.
 The symmetric tunnel-coupling between the dots and the leads 
 will be assumed to be independent of energy, $|V_{k}| = |V|$.

 The current flowing into the each quantum dot
 can be defined as the rate of change in the number of electrons in a lead.
 At the left lead $L$, the total current is split into two
 local currents, $I_1$ and $I_2$.
 The commutator of the number operator
 $N = \sum_{k\sigma \in L} c^\dagger_{k\sigma} c_{k\sigma} $
 with the Hamiltonian (\ref{Hamiltonian}) gives rise to the current
 as the sum of the local currents through each dot,
\begin{mathletters}
\begin{eqnarray}
I &=& \sum_{i=1,2} I_i, \\
I_i &= & -\frac{e}{h} {\mathrm Re} \Bigl\lbrace \sum_{k\sigma \in L}
   \int d\varepsilon \, V_{k}\,G^<_{k\sigma,i\sigma}(\varepsilon) \Bigr\rbrace 
\end{eqnarray}
\end{mathletters}
 with the nonequilibrium Green's function 
 $G^<_{k\sigma,i\sigma}(t-t') \equiv i\langle d^\dagger_{i\sigma}(t') 
                               c_{k\sigma}(t) \rangle$.
 With the Keldysh technique
 for nonlinear current through the system,
 the local currents through each dot are given by
 \cite{Meir92,Cho99}
\begin{equation}
   I_i = {e\over h} \sum_\sigma \int d \varepsilon \, 
       (f_L(\varepsilon)-f_R(\varepsilon)) {\cal T}_i(\varepsilon),
\end{equation}
 where the local transmission spectral functions are defined by
   ${\cal T}_i(\varepsilon)= 
     \left\{{\bf \Gamma}^L {\bf G}^{\rm r}_\sigma (\varepsilon)
     {\bf \Gamma}^R {\bf G}^{\rm a}_\sigma (\varepsilon) \right\}_{ii} $
 which is the $i$-th diagonal component of the matrix transmission
 spectral function.
 Here, $f_\alpha (\varepsilon)=f(\varepsilon-\mu_\alpha)$ 
 is the Fermi-Dirac distribution function of the leads
 $\alpha=L,R$ and $\mu_L=-\mu_R=eV/2$ with applied bias $eV$ between two leads.
 Due to tunneling each dot level acquires a finite line width
 $\Gamma=2\pi|V|^2 {\cal N}$,
 where ${\cal N}$ is the density of states in the leads.
 The matrix coupling to the leads is described by
 ${\bf \Gamma}^L={\bf \Gamma}^R=
 \Gamma \left(\begin{array}{cc} 1 & 1 \\ 1 & 1 \end{array} \right)$.
 ${\bf G}^{\rm r}_\sigma (\varepsilon)$ 
 is the matrix dot Green's function defined in time space as 
 $G_{ij,\sigma}^{\rm r}(t-t')
 =-i\theta (t-t')\langle \{ d_{i\sigma}(t),d_{j\sigma}^\dagger(t')\}\rangle$.
 By using the equation of motion treatment,
 one can obtain the matrix Green's function of the dots as
\begin{equation}
\label{Green's function}
   {\bf G}_\sigma^{\rm r}(\varepsilon) =
        \left( \begin{array}{cc}
        \varepsilon - \varepsilon_1 + i \Gamma & i \Gamma \\
        i \Gamma & \varepsilon - \varepsilon_2 + i \Gamma
        \end{array} \right) ^{-1}
\end{equation}
 and ${\bf G}_\sigma^{\rm a}(\varepsilon)
      =[{\bf G}_\sigma^{\rm r}(\varepsilon)]^\dagger$.
 Accordingly, the local transmission spectral functions are written by
\begin{mathletters}
\begin{eqnarray}
  {\cal T}_1(\varepsilon) &=& \!\!
  {\Gamma^2 (\varepsilon -\varepsilon_2) 
  (2\varepsilon - \varepsilon_1-\varepsilon_2) 
   \over
  (\varepsilon - \varepsilon_1)^2 (\varepsilon-\varepsilon_2)^2 
    + (2\varepsilon - \varepsilon_1-\varepsilon_2)^2 \Gamma^2 }\, , \\
  {\cal T}_2(\varepsilon) &=& \!\!
  {\Gamma^2 (\varepsilon -\varepsilon_1) 
  (2\varepsilon - \varepsilon_1-\varepsilon_2) 
   \over
  (\varepsilon - \varepsilon_1)^2 (\varepsilon-\varepsilon_2)^2 
    + (2\varepsilon - \varepsilon_1-\varepsilon_2)^2 \Gamma^2 }.
\end{eqnarray}
\end{mathletters}
 Note that these can be negative.
 The total current is the sum of current through each dot 
 $I=I_1+I_2$ which is just the current conservation.
 This leads to the total transmission spectral function as
 ${\cal T}(\varepsilon)={\cal T}_1(\varepsilon)+{\cal T}_2(\varepsilon)$,
\begin{equation}
  {\cal T}(\varepsilon)  =  
  {\Gamma^2 (2\varepsilon - \varepsilon_1-\varepsilon_2)^2 \over
     (\varepsilon - \varepsilon_1)^2 (\varepsilon-\varepsilon_2)^2 
      + (2\varepsilon - \varepsilon_1-\varepsilon_2)^2 \Gamma^2 }\, . 
\end{equation}
 We note that this is always positive.
 The classical analogue of our system is two resistors in parallel.
 $I_1$ and $I_2$ must then both be positive.
 In contrast, in a quantum system the only constraint is that
 current conservation requires $I=I_1+I_2$.
 It is not required that $I > I_1$, $I_2$.
 For the case of a metallic ring coupled to leads, 
 this was pointed out previously by
 Jayannavar and Deo\cite{Jayan95}

 Let us assume the cases of $I < I_1$ or $I < I_2$ 
 under the current conservation for $\mu_R < \mu_L$.
 For given energy levels $(\varepsilon_1,\varepsilon_2)$,
 if $I(\varepsilon_1,\varepsilon_2) < I_1(\varepsilon_1,\varepsilon_2)$,
 we can assign an excess current $I_{exc}(\varepsilon_1,\varepsilon_2)$.
 Then we can rewrite the total current as
 $I(\varepsilon_1,\varepsilon_2)=I_1(\varepsilon_1,\varepsilon_2)
 -I_{exc}(\varepsilon_1,\varepsilon_2)$.
 The current conservation gives rise to the local excess current 
 of $I_{exc}(\varepsilon_1,\varepsilon_2) =-I_2(\varepsilon_1,\varepsilon_2)$
 which should circulate {\em clockwise} on the closed path through
 the dots and the leads. 
 In the opposite case of $I < I_2$, the excess current becomes
 $I_{exc}(\varepsilon_1,\varepsilon_2) = -I_1(\varepsilon_1,\varepsilon_2)$
 circulating {\em counter-clockwise} on the closed path.
 The circulating current makes the device magnetically polarized.
 Therefore, we define the circulating current as a magnetic polarization
 current (MPC) $I_M \equiv -I_{exc}$. 
 We choose its direction for the case of $I < I_1$ as {\em positive}.
 It should be noted that this is purely
 a quantum coherent mesoscopic phenomena.

 Considering the transport current (TC), $I$, and 
 MPC, 
 $I_M$, on an equal footing,
 we define the MPC as
\begin{equation}
 I_M = \frac{-e}{h} \sum_\sigma \int d\varepsilon
     (f_L(\varepsilon)-f_R(\varepsilon)) {\cal T}_M(\varepsilon)  
 \label{MPC}
\end{equation}
 with the effective spectral function, ${\cal T}_M(\varepsilon)$.
 The ${\cal T}_M(\varepsilon)$ can be extracted from the following arguments.
 Let us recall the transmission spectral functions 
 for $\varepsilon_1 < \varepsilon_2$.
 ${\cal T}(\varepsilon)$ has three extremum points,
 that is, ${\cal T}(\varepsilon_1)= {\cal T}(\varepsilon_2)=1$
 (resonant transmission)
 and ${\cal T}(\bar \varepsilon)=0$
 (anti-resonant transmission),
 where ${\bar \varepsilon}=(\varepsilon_1+\varepsilon_2)/2$.
 At $\varepsilon = {\bar \varepsilon}$,
 the anti-resonance of ${\cal T}(\varepsilon)$ gives rise to
 a pronounced dip structure originating from
 the destructive interference between the transmissions through one QD
 and the other.
 Such an anti-resonant feature in a transport system with
 two different transmission channels is well understood
 as the Fano effect\cite{Fano}.
 Next, the two local transmission spectrals of 
 ${\cal T}_1(\varepsilon)$ and ${\cal T}_2(\varepsilon)$
 have three characteristic points, that is,
 ${\cal T}_1(\varepsilon_1)=1 \left[ {\cal T}_2(\varepsilon_2)= 1 \right]$
 and 
 ${\cal T}_1(\varepsilon_2)={\cal T}_1({\bar \varepsilon})=0$ 
 $\left[
 {\cal T}_2(\varepsilon_1)= {\cal T}_2({\bar \varepsilon})=0 \right]$.
 These points have nothing to do with resonant and anti-resonant
 tunneling through each dot.
 The two local transmission spectral functions
 only give us 
 information about the local currents.
 Then we have to determine the behavior of the local spectrals
 in other energy regions.
 It is convenient to consider the ratio of the local transmission
 to the total transmission.
 The ratios are written by
 ${\cal T}_1(\varepsilon)/{\cal T}(\varepsilon)= 1/(1+g(\varepsilon))$ and 
 ${\cal T}_2(\varepsilon)/{\cal T}(\varepsilon)= 1/(1+g(\varepsilon)^{-1})$,
 where 
 $g(\varepsilon)=(\varepsilon-\varepsilon_1)/(\varepsilon-\varepsilon_2)$.
 For $\varepsilon < \varepsilon_1$ and $\varepsilon > \varepsilon_2 $,
 since $0< g(\varepsilon) <1$, the ratios are between 0 and 1.
 In these regions, there are no local excess currents.
 However,
 for $\varepsilon_1 < \varepsilon < {\bar \varepsilon}$,
 ${\cal T}_1(\varepsilon)/{\cal T}(\varepsilon) > 1$
 $\left[ {\cal T}_2(\varepsilon)/{\cal T}(\varepsilon) < 0 \right]$
 and
 for ${\bar \varepsilon} < \varepsilon < \varepsilon_2$, 
 ${\cal T}_1(\varepsilon)/{\cal T}(\varepsilon) < 0$
 $\left[ {\cal T}_2(\varepsilon)/{\cal T}(\varepsilon) > 1 \right]$.
 These spectral properties give rise to the MPC at a given energy $\varepsilon$.
 Similarly, one can decide the ${\cal T}_M(\varepsilon)$ for 
 $\varepsilon_1 > \varepsilon_2$.
 Consequently, we obtain the ${\cal T}_M(\varepsilon)$ in terms of
 the local transmission spectrals as
\begin{eqnarray}
 {\cal T}_M(\varepsilon) 
  &=& \sum_{i \neq j} 
      \theta(\varepsilon_i - \varepsilon_j) 
   \Big\{
      \theta(\varepsilon - {\bar \varepsilon} )
      \theta(\varepsilon_i - \varepsilon) {\cal T}_j(\varepsilon) 
   \Big. \nonumber \\ &&  \hspace{2.2cm}  \Big.
   -  \theta(\varepsilon - \varepsilon_j)
      \theta({\bar \varepsilon} - \varepsilon) 
      {\cal T}_i(\varepsilon) \Big\} \, .
\label{MPC_spectral}
\end{eqnarray}
 Equations (\ref{MPC}) and (\ref{MPC_spectral}) 
 are the central result of this work.
 Note that, for a given energy level position 
 $(\varepsilon_1,\varepsilon_2)$ in each dot,
 the ${\cal T}_M(\varepsilon)$ is non-zero
 between the two energy levels $\varepsilon_i < \varepsilon < \varepsilon_j$ 
 and is an anti-symmetric function with respect to $\bar \varepsilon$,
 ${\cal T}_M(\varepsilon-\bar \varepsilon)
  =-{\cal T}_M(\bar \varepsilon-\varepsilon)$.
 These properties of ${\cal T}_M(\varepsilon)$ determine
 the window of applied bias in which the MPC can be measured. 

\begin{figure}
\vspace*{5cm}
\includegraphics{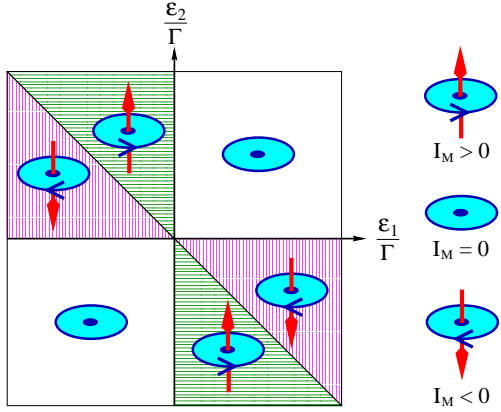}
\caption{
 Magnetic polarization of the double quantum dot device
 as a function of the energy level positions
 of each dot $(\varepsilon_1/\Gamma,\varepsilon_2/\Gamma)$,
 in the limit of zero applied bias at zero temperature
 for $\mu_R < \mu_L$.
 The vertical arrows stand for the magnetic moment of the DQD device
 whose length and direction depends on the amplitude 
 and direction of the magnetic polarization current, respectively.
 }
\label{fig2}
\end{figure}

 At zero temperature,
 the limit of zero applied bias is the most simple case.
 The TC is proportional to the 
 transmission of incoming electrons at the Fermi energy ($\varepsilon_F=0$);
 $\lim_{eV \rightarrow 0} I 
 = (2e/h) {\cal T}(\varepsilon)|_{\varepsilon=\varepsilon_F}eV$ 
 and the MPC becomes
 $\lim_{eV \rightarrow 0} I_M 
 = (-2e/h) {\cal T}_M(\varepsilon)|_{\varepsilon=\varepsilon_F}eV$.
 When the energy level of one dot is lying below the Fermi energy
 and that of the other is lying above the Fermi energy,
 the MPC appears to polarize the DQD device.
 If both energy levels of dots are below or above the Fermi energy
 the device is not magnetically polarized.
 This implies that the interference between the electron and hole channels
 produces the MPC.
 Figure \ref{fig2} shows 
 the magnetic polarization as a function 
 of $(\varepsilon_1/\Gamma,\varepsilon_2/\Gamma)$
 for $\mu_R < \mu_L$.
 It is shown that manipulating the energy level position of each dot,
 one can magnetize the DQD devices as {\em up}-, {\em non}-, or 
 {\em down}-polarized.
 Applying a finite bias between the leads,
 the properties of ${\cal T}_M(\varepsilon)$ change the polarization zone
 boundaries.
 A finite applied bias develops a {\em non}-polarization zone 
 satisfying the conditions of $-eV/2 < |\varepsilon_1-\varepsilon_2| < eV/2$
 or $-eV/2 < \varepsilon_1,\varepsilon_2 < eV/2$.
 While the {\em up}-, and {\em down}-polarization zones are extended to
 the {\em non}-polarization zone of the limit of zero applied bias
 because the electron and hole channels near the Fermi energy
 within the window of the applied bias contribute to the MPC. 
 It should be noted,
 when the applied bias is reversed to $\mu_R > \mu_L$,
 the magnetic moment of the device is reversed.

\begin{figure}
 \vspace*{10.0cm}
 \includegraphics{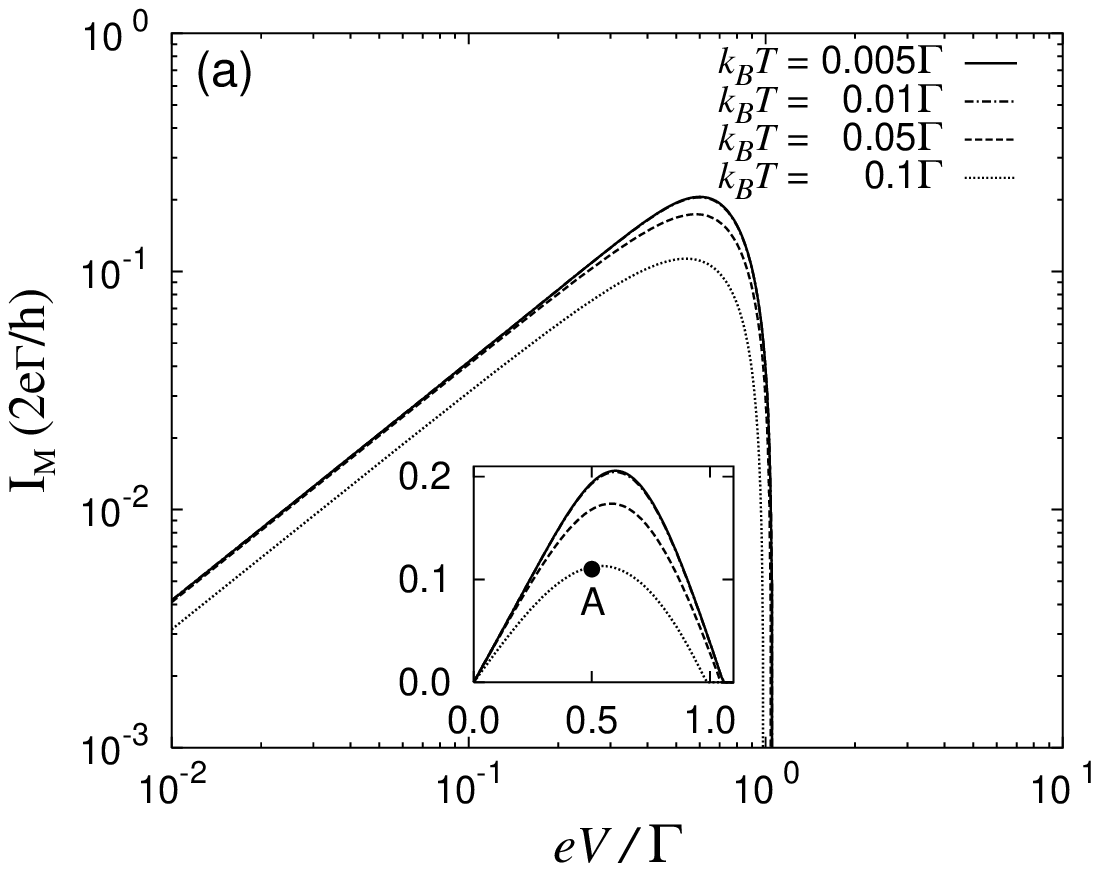}
 \includegraphics{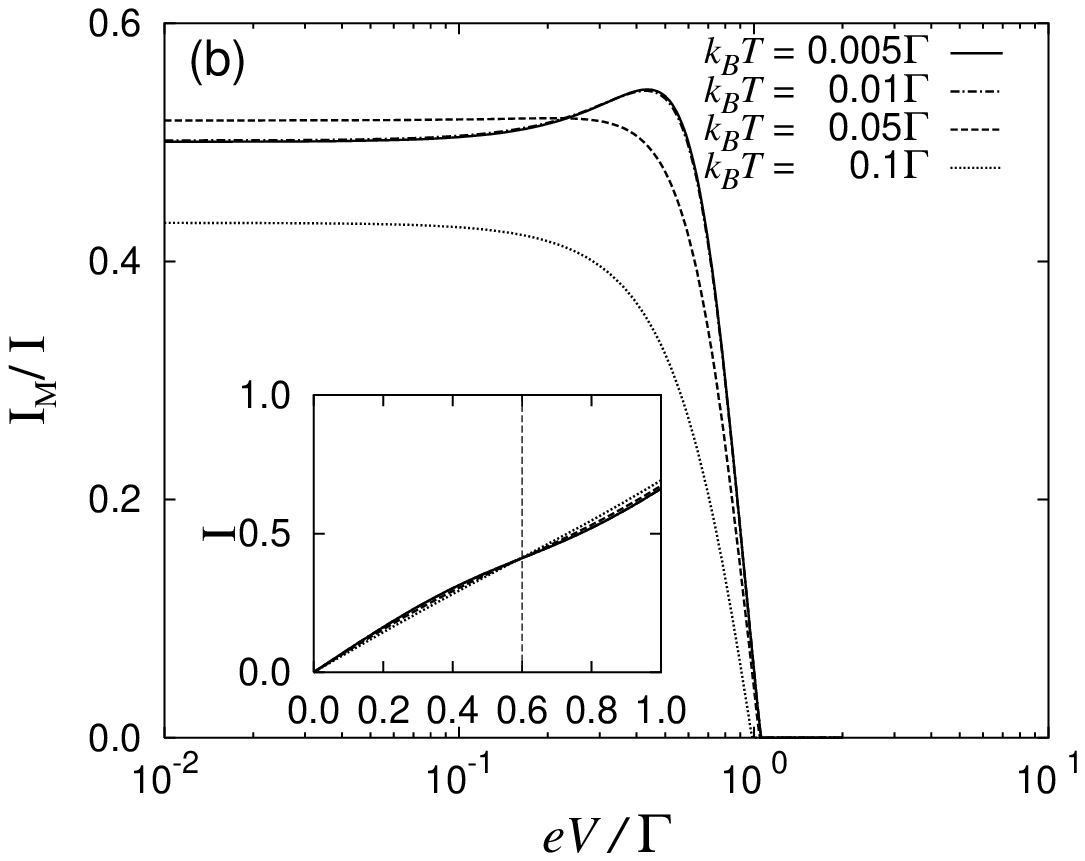}
\caption{
 (a) Magnetic polarization current 
 and 
 (b) its ratio to the transport current 
 as a function of applied bias
 at $(\varepsilon_1/\Gamma,\varepsilon_2/\Gamma)=(0.3,-0.9)$
 for various temperatures  $k_BT$.
 In the insets, the currents are shown in a linear scale.
 }
\label{fig3}
\end{figure}

\begin{figure}
\vspace*{10.0cm}
\includegraphics{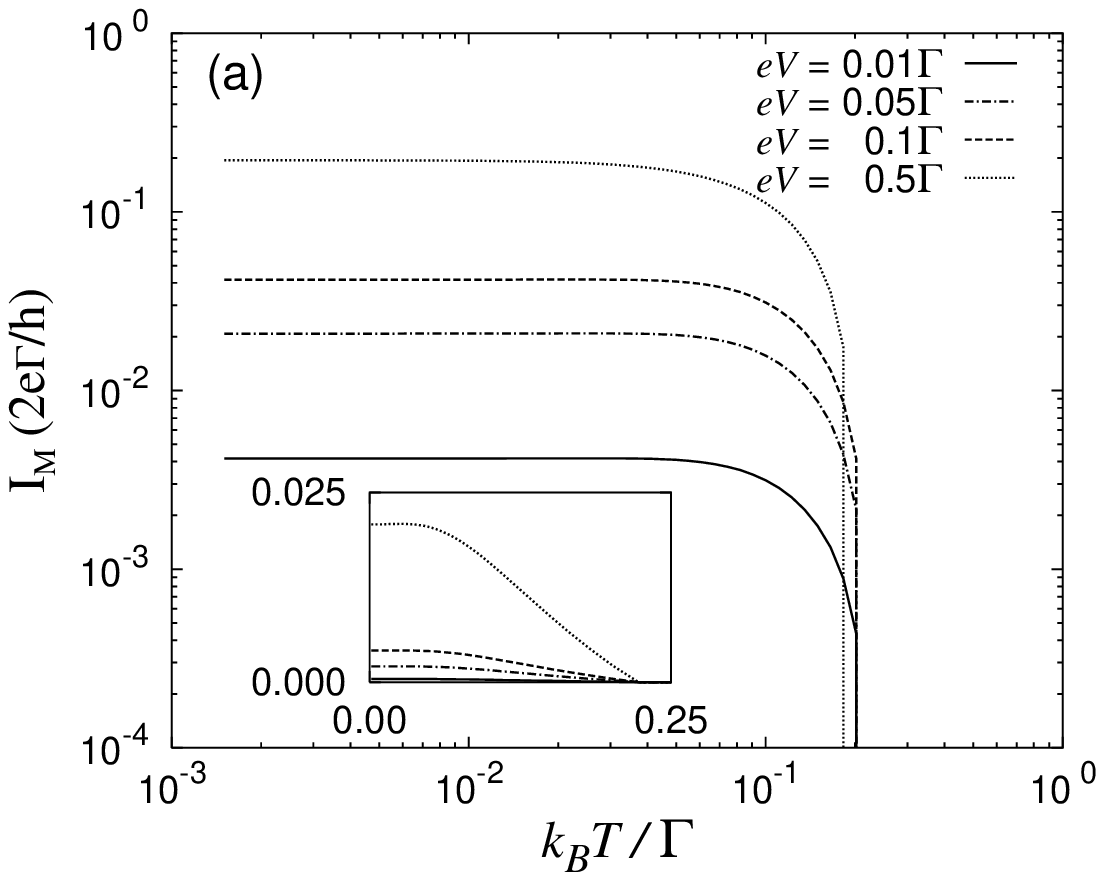}
\includegraphics{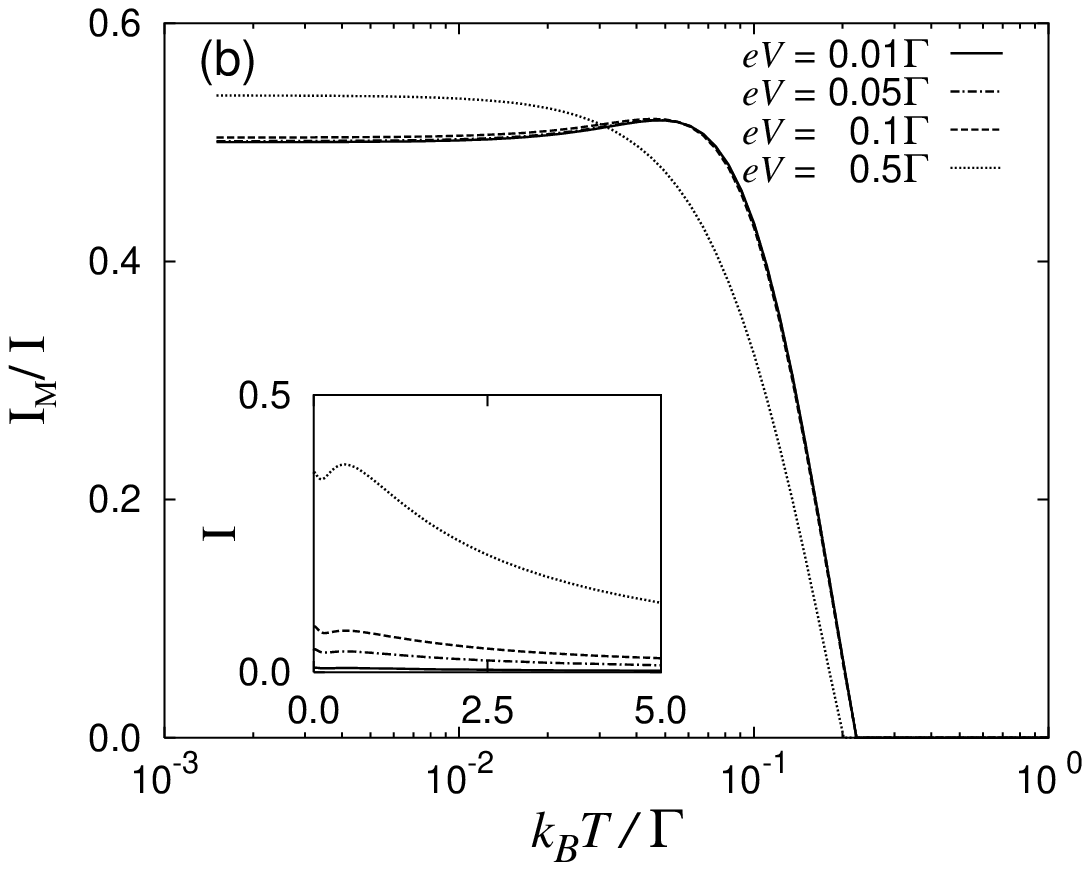}
\caption{Temperature dependence of (a) the magnetic polarization current 
 and (b) its ratio to the transport current 
 at $(\varepsilon_1/\Gamma,\varepsilon_2/\Gamma)=(0.3,-0.9)$
 for different values of the applied bias $eV$.
 In the insets, the currents 
 are shown in a linear scale.
 }
\label{fig4}
\end{figure}

 To illustrate the MPC for finite temperature,
 we choose a set of energy level position 
 $(\varepsilon_1/\Gamma,\varepsilon_2/\Gamma)=(0.3,-0.9)$
 which can be adjusted to other values by varying the gate voltages.
 In fact, the level positions taken in the same polarization zone
 do not affect the physics
 of the MPC but only change its amplitude.
 We display the MPC and the TC as a function of applied bias 
 for different temperatures in Fig. \ref{fig3}.
 As the applied bias increases from zero bias,
 both the TC and the MPC increase linearly.
 The MPC is always smaller than the TC
 for these given energy level positions. 
 However, for the case of other energy level positions,
 the MPC can become larger than the TC
 (e.g., for $(\varepsilon_1/\Gamma,\varepsilon_2/\Gamma)
 =(0.5,-0.6)$, $I_M \simeq 5 \, I $ at low temperatures).
 This linear behavior of the MPC shows 
 that the MPC emerges only in nonequilibrium.
 Further increase of applied bias results in the MPC
 approaching its maximum value. 
 Eventually, 
 the disappearance of the MPC arises
 when the window of the applied bias becomes larger than
 the range of $\varepsilon$ for which
 ${\cal T}_M(\varepsilon)$
 has a non-zero value $(-0.9 < \varepsilon < 0.3)$.
 The inset of Fig. \ref{fig3}(b) shows that
 the TC increases non-linearly as the applied bias increases.
 In addition, compared to the MPC,
 the TC is suppressed for the bias smaller
 than $eV=0.6\, \Gamma$ but enhanced for the bias larger
 than $eV=0.6\, \Gamma$ by thermal effects.  
 This originates from the fact that
 ${\cal T}(\varepsilon)$ has a pronounced dip structure 
 at $\bar \varepsilon = -0.3\, \Gamma$, due to the Fano effect.
 However, the anti-symmetric property of ${\cal T}_M(\varepsilon)$
 gives rise to moderate thermal suppression of the MPC.
 The thermal suppression is manifestly shown
 in the temperature dependence of the MPC and the TC in Fig. \ref{fig4}.
 The relatively large applied bias leads to the large amplitude of the MPC.
 This is consistent with 
 the linear behavior of the MPC in the $I_M$-$V$ curve.
 The insets of Fig. \ref{fig4}(a) and (b) show,
 compared to the TC, the more rapid suppression of the MPC
 since ${\cal T}_M(\varepsilon)$ 
 is zero for low and high energies.
 At $k_BT \simeq 0.05 \, \Gamma$,
 the MPCs at various applied biases start to be suppressed by 
 thermal effects. 
 For temperatures higher than $k_BT \simeq 0.2 \, \Gamma$,
 thermal effects wash out this novel quantum
 coherent phenomena.
 
 From the experimental parameters measured in Ref.\cite{Holleitner00};
 $\Gamma \simeq 50 \, \mu{\rm eV}$ and 
 ${\cal A} = 2.52 \times 10^{-13} {\rm m}^2$,
 where ${\cal A}$ is a corresponding area to AB oscillation,
 we can estimate the amplitude of a MPC and an induced magnetic moment,
 $|\vec \mu_D|= {\cal A} \cdot I_M$.
 At the point {\sf A} in the inset of Fig. \ref{fig3}(a),
 for $k_BT=0.1\, \Gamma \, (T \simeq 50 \, {\rm mK})$,
 one can estimate $I_M \simeq 0.36 \, {\rm nA}$,
 when $eV \simeq 25 \mu{\rm eV}$ is applied.
 The induced magnetic moment of the device becomes 
 $|\vec \mu_D| \simeq 9\, \mu_B$, where $\mu_B$ is the Bohr magneton. 
 Comparing this estimate of
 the MPC to recent measurements of persistent currents
 \cite{Rabaud01}  
 suggests that the effects we are discussing can be
 observed with existing nanotechnology.

%
 In summary, 
 we studied coherent electron transport through two parallel quantum dots,
 each of which has a single tunable energy level.
 By changing these energy levels
 in the DQD device one can vary the
 {\it sign} and {\it magnitude} of the 
 magnetic polarization current induced by
 quantum interference effects.
 This current is sufficiently large that
 it should be experimentally observable. 

This work was supported by the University of Queensland,
the Australian Research Council,
and
the Korean Science and Engineering
Foundation through the Center for Strongly Correlated Materials Research (SNU).


\end{document}